\documentclass[12pt,preprint]{aastex}

\usepackage{graphics,graphicx}
\usepackage{subfigure}
\usepackage{color}
\usepackage{lscape}

\begin{document}

\title{On the Star Formation-AGN Connection at $z \lesssim 0.3$}

\author{Stephanie M. LaMassa$^{1}$, T. M. Heckman$^{2}$, A. Ptak$^{3}$, C. Megan Urry$^{1}$}

\affil{$^1$Yale University - Department of Physics; Yale Center for Astronomy \& Astrophysics
$^2$The Johns Hopkins University
$^3$ NASA/GSFC
}

\begin{abstract}
Using the spectra of a sample of $\sim$28,000 nearby obscured active galaxies from Data Release 7 of the Sloan Digital Sky Survey (SDSS), we probe the connection between AGN activity and star formation over a range of radial scales in the host galaxy. We use the extinction-corrected luminosity of the [OIII] 5007 \AA\ line as a proxy of intrinsic AGN power and supermassive black hole (SMBH) accretion rate. The star formation rates (SFRs) are taken from the MPA-JHU value-added catalog and are measured through the 3$^{\prime\prime}$ SDSS aperture. We construct matched samples of galaxies covering a range in redshifts. With increasing redshift, the projected aperture size encompasses increasing amounts of the host galaxy. This allows us to trace the radial distribution of star-formation as a function of AGN luminosity. We find that the star formation becomes more centrally concentrated with increasing AGN luminosity and Eddington ratio. This implies that such circumnuclear star formation is associated with AGN activity, and that it increasingly dominates over omnipresent disk star formation at higher AGN luminosities, placing critical constraints on theoretical models that link host galaxy star formation and SMBH fueling. We parametrize this relationship and find that the star formation on radial scales $<$1.7 kpc, when including a constant disk component, has a sub-linear dependence on SMBH accretion rate: $SFR \propto \dot{M}^{0.36}$, suggesting that angular momentum transfer through the disk limits accretion efficiency rather than the supply from stellar mass loss.
\end{abstract}

\section{Introduction}
Supermassive black holes (SMBHs) are intrinsically linked to their parent galaxies. As the black hole grows via accretion in the Active Galactic Nucleus (AGN) phase, the bulge of the host galaxy builds up from star formation, as evidenced by the correlation between SMBH mass and velocity dispersion of the host galaxy \citep{kr, magorrian, FM, gebhardt, tremaine, hr}. Although the evolution of SMBH accretion and star formation are surely intertwined, the mechanisms that couple these processes are not definitively known. Gas can be driven to the potential well of the black hole by galaxy mergers, which compress gas and thereby trigger host galaxy star formation \citep[e.g.,][]{Sanders, kh, dimatteo, springel, hopkins05}. Alternatively, secular mechanisms such as torques from galactic bars, bars-within-bars, spiral arms \citep[][]{sfb, gb, hopkins10}, stochastic accretion of cold gas \citep{hopkins06} or mass loss from stellar winds \citep{ns, co97, co07, kh09} can contribute significantly to both processes. Feedback is another key ingredient in the star formation and AGN connection, where AGN winds can blow out circumnuclear gas, suppressing star formation and regulating SMBH and bulge growth \citep{silk,  haehnelt, granato, dimatteo, wyithe, somerville}. Observational studies of local optically obscured AGN (Type 2 Seyferts, Sy2s) have indicated that star formation processes can both feed the black hole, via stellar mass loss, and limit accretion efficiency through supernova feedback \citep{wild}.

Seyfert 2 galaxies are prime candidates for studying the observational link between SMBH growth and host galaxy star formation. As the accretion disk is enshrouded by circumnuclear dust obscuration, optical signatures of star formation in the host galaxy as well as AGN fueling can be investigated. Previous studies of Sy2s have revealed that a significant fraction live in host galaxies that are undergoing (or have recently experienced) significant amounts of star formation \citep{kauff, cf04, netzer}. Using a sample of over 20,000 AGN from the Sloan Digital Sky Survey (SDSS) Data Release One \citep{DR1}, \citet{kauff} demonstrated that the trend between star formation activity and AGN accretion was not confined to the nuclear regions of the host galaxies for luminous AGN. Subsequently, \citet{k07} used a combination of SDSS spectra and GALEX UV images to show that the link between star-formation and black hole growth was much more direct for star formation in the central few kpc than for star formation in the global galactic disk. In agreement with this result, \citet{ds} studied 74 galaxies from the revised Shapley-Ames Seyfert sample \citep{rsa, mr, ho} and found that this connection is strongest on sub-kpc scales, with a much weaker correlation when considering larger scale (i.e., $>$1 kpc) star formation. These place important constraints on the mechanisms by which star formation and black hole growth are linked.

However, these studies suffer from non-uniform methodology for measuring star formation rates and/or small sample sizes. To address these limitations, in this work we test the relationship between star formation and SMBH accretion on various galactic scales using a sample of $\sim$28,000 type 2 AGN and composite galaxies culled from SDSS. All quantities are measured in a uniform way through the 3$^{\prime\prime}$ spectroscopic fiber. These quantities parametrize the AGN luminosity, SMBH accretion rate, and star formation rate (SFR). We test whether the connection between AGN fueling and starburst activity varies as the projected size of the SDSS aperture sequentially covers greater parts of the host galaxy, and whether this relationship is a function of intrinsic AGN luminosity.

\section{Data Analysis}
Our data were drawn from the Main Galaxy Sample of SDSS Data Release 7 \citep[DR7][]{Strauss, dr7} where we utilized the MPA-JHU value added catalogs.\footnote{http://www.sdss3.org/dr9/algorithms/galaxy\_mpa\_jhu.php} We isolated sources categorized as ``AGN'' and ``Composite'' according to the diagnostic BPT diagram \citep{bpt, kewley, kauff}; we note these classifications were made automatically by the MPA-JHU pipeline. We exclude galaxies classified as LINERs. We imposed several quality control cuts: the H$\alpha$, H$\beta$, [OIII] 5007\AA\ and [NII] 6584 \AA\ emission lines had to be significant above the 5$\sigma$ level and the H$\alpha$ equivalent width had to exceed 3 \AA\, removing galaxies where the ionizing radiation may be due to low-mass evolved stars and not nuclear accretion \citep{CF}. We also restrict the sample to the Petrosian magnitude range 15.5 $< m_r <$ 17.8 (the limiting magnitude of the Main Galaxy Sample) and remove outliers in BPT parameter space, focusing between -1.0 $<$ log([NII]/H$\alpha$) $<$ 1.0 and -1.0 $<$ log([OIII]/H$\beta$) $<$ 2. The objects in our sample span 0.01 $< z < 0.32$. As shown by \citet{kauff} and \citet{Reichard}, these AGN host galaxies are typically structurally normal galaxies of intermediate Hubble type (average SDSS concentration index of 2.61$\pm$0.33).

\subsection{AGN Activity}
\begin{figure}[ht]
{\includegraphics[scale=0.8,angle=90]{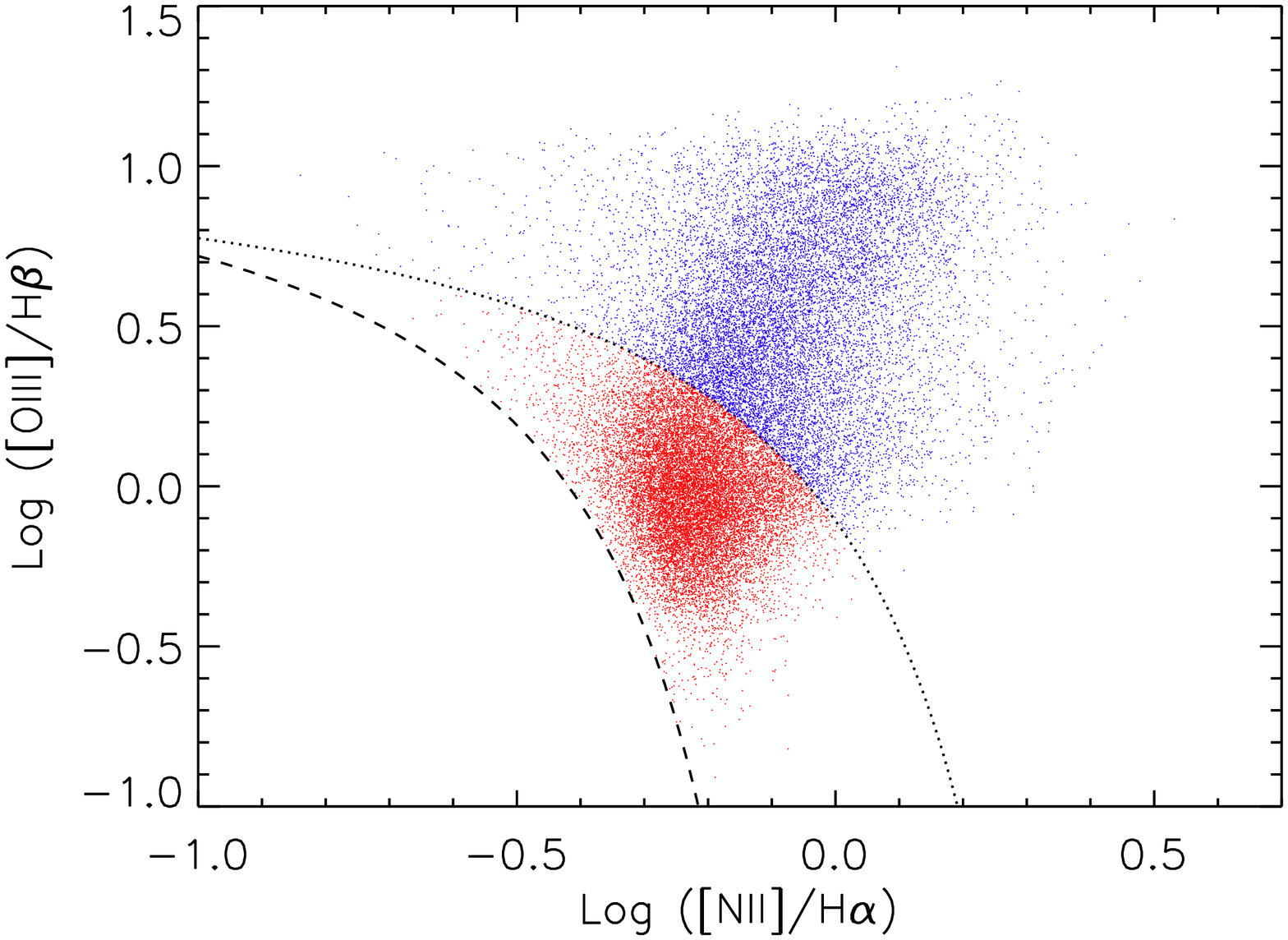}}
\caption[]{\label{our_bpt} BPT diagram for galaxies used in our analysis. The lower dashed line represents the empirical boundary between star forming galaxies and active systems from \citet{kauff} while the upper dotted line delineates the maximum theoretical starburst line from \citet{kewley}. There are 15508 composite galaxies (red data points, between the two curves) and 12444 Sy2 galaxies (blue data points). These composite systems represent the subset of composites in which the AGN contribution to [OIII] emission is estimated at $\geq$ 30\% of the total.}
\end{figure}

We use the luminosity of the [OIII] line (L$_{[OIII]}$) as a proxy for intrinsic AGN luminosity since it is primarily ionized by accretion disk photons. Since [OIII] forms in the AGN narrow line region, it is not affected strongly by dust obscuration from the circumnuclear torus, providing a good estimate of the intrinsic AGN luminosity in SDSS galaxies \citep{kauff, heckman, me}. For composite galaxies, i.e., those systems between the \citet{kauff} empirical boundary between star-forming and active galaxies and the \citet{kewley} theoretical maximum starburst line on the BPT diagram, we estimated the AGN contribution to the [OIII] emission following the prescription of \citet{wild}. This method estimates the star formation contributions to L$_{[OIII]}$ based on the location of the source in the BPT diagram. We only keep composite galaxies if the AGN component of the [OIII] line is at least 30\% of the total [OIII] emission, removing very weak AGN. In total, we have 27952 galaxies, of which 15508 are composites and 12444 are Sy2s; they are shown on the BPT diagram in Figure \ref{our_bpt}. Finally, we correct L$_{[OIII]}$ for dust obscuration using the observed Balmer decrement (H$\alpha$/H$\beta$), assuming an intrinsic H$\alpha$/H$\beta$ value of 3.1, and the standard $R(V)=A(V)/E(B-V)$=3.1 extinction curve \citep{Cardelli}, obtaining L$_{[OIII],corr}$.

The SMBH mas (M$_{BH}$) goes as the host galaxy stellar velocity dispersion ($\sigma$) to the fourth power \citep{tremaine}. To parametrize the SMBH accretion rate, we therefore define a pseudo-Eddington parameter, L$_{[OIII],corr}$/$\sigma^4$, for the galaxies where a velocity dispersion was measured. In the following, we search for trends between AGN luminosity (L$_{[OIII],corr}$) and SFR, and between the Eddington parameter (L$_{[OIII],corr}$/$\sigma^4$) and specific SFR (sSFR, the SFR normalized by galaxy mass).

\subsection{Star Formation}
We focus on the SFR and sSFR measured through the 3$^{\prime\prime}$ SDSS spectroscopic fiber (SFR$_{fiber}$, sSFR$_{fiber}$), calculated as outlined in \citet{Brinchmann}. As nuclear activity contributes to H$\alpha$ luminosity, which is used to calibrate the SFR in quiescent galaxies, these SFRs are instead calculated using the break in the optical spectrum at 4000 \AA\ \citep[D$_{n}$(4000),][]{Balogh}, which is a measure of the age of the host galaxy stellar population: a shallower break is associated with a younger stellar population. Using the relationship between Dn(4000) and the H$\alpha$ calibrated sSFR$_{fiber}$ for star-forming galaxies, \citet{Brinchmann} constructed probability distribution functions from which sSFR$_{fiber}$ were estimated for active galaxies; SFR$_{fiber}$ is then calculated by multiplying sSFR$_{fiber}$ by the galaxy mass within the fiber. In this analysis, we use the value-added catalog provided median estimates of the SFR$_{fiber}$ and sSFR$_{fiber}$ probability distribution functions.

\subsection{$z/z_{max}$ as a Distance Proxy}
Our goal is to determine the relationship between black hole growth and star formation over a range of physical scales inside the galaxy. To ensure that we are comparing host galaxies with similar properties at all distances, and isolating the physical size of the projected SDSS fiber aperture as the varying parameter, we introduce the $z/z_{max}$ parameter \citep{kauff,kauff2}, where $z_{max}$ is the redshift at which the galaxy falls out of the apparent magnitude limit of the sample ($m_r = 17.8$). Absent significant cosmic evolution (see below), the properties of a population of galaxies selected from SDSS will be invariant with $z/z_{max}$. Normalizing the redshift of each galaxy by $z_{max}$ corrects for the effect that more luminous and massive galaxies are visible out to greater distances. To calculate $z_{max}$, we first calculate the absolute $r$ magnitude ($M_r$) using:
\begin{equation}
M_R = m_r - DM(z) - k(z),
\end{equation}

where $DM(z)$ is the distance modulus and $k(z)$ is the k-correction. We adopted a cosmology of H$_0$ = 70 km s$^{-1}$ Mpc$^{-1}$, $\Omega_M$= 0.27 and $\Lambda$=0.73.

We determined k-corrections for each galaxy using {\it kcorrect} version 4.2 \citep{kcorr}. Using $M_R$, we then found the redshift, $z_{max}$, at which each galaxy would have an apparent magnitude of 17.8. As indicated in the legends of Figure \ref{sfr_agn}, $z/z_{max}$ is proportional to redshift. 

\subsection{The Effects of Cosmic Evolution \& Sample Completeness}
With our approach of using galaxies over a range of redshifts to study the radial distribution of star-formation in AGN, there are two issues that must be addressed. Firstly, since we are confining our study to the relatively local universe, we do not expect the number density of AGN to evolve with redshift. Our sample binned in $z/z_{max}$ only span a range in median redshift from 0.06 to 0.13. Over this range, the increase in the rate of star-formation and black hole growth per co-moving volume element (e.g. Marconi et al. 2004) is only a factor of $\sim$ 1.2 (0.08 dex). Secondly, low-luminosity AGN will be increasingly difficult to recognize in more distant galaxies since the diagnostic emission-lines will become fainter and more diluted by emission due to host galaxy star formation.

To address these two issues, we performed a $V/V_{max}$ test for our sample, where $V$ is the cosmic volume probed by an individual galaxy sample and $V_{max}$ is the maximum volume out to which that galaxy would be observable before dropping below the sensitivity limit of the survey. Values for the median of a $V/V_{max}$ distribution of significantly less than (greater than) 0.5 signals incompleteness (cosmic evolution). In Table \ref{v_vmax}, we list the median $V/V_{max}$ for various AGN luminosity and Eddington parameter bins. In the following, we work within the range where $V/V_{max} = 0.5 \pm 0.1$, i.e., L$_{[OIII],corr} \geq 10^{40}$ erg s$^{-1}$ and L$_{[OIII],corr}$/$\sigma^{4} \geq 3\times10^{31}$ erg s$^{3}$ km$^{-4}$. We also find that the median estimate of the total stellar mass and the velocity dispersion does not evolve with $z/z_{max}$ in these AGN luminosity and Eddington parameter bins. Thus, we are comparing similar galaxies in aggregate over increasing $z/z_{max}$ values.

Adopting the bolometric correction for the extinction-corrected [OIII] luminosity from \citet{kh09} and the conversion from $\sigma$ to M$_{BH}$ from \citet{tremaine}, our sample covers ranges from roughly $10^{9.2}$ to $10^{12.2}$ L$_{\odot}$ in bolometric luminosity and from roughly $10^{-2.8}$ to 1 in $L/L_{Edd}$. Over these broad ranges we are reasonably complete and the effects of cosmic evolution are modest.

\begin{deluxetable}{llll}

\tablewidth{0pt}
\tablecaption{\label{v_vmax} Median $V/V_{max}$ for AGN Luminosity and Eddington Parameter Bins}
\tablehead{

\colhead{Log (L$_{[OIII],corr}$)} & \colhead{$V/V_{max}$} & \colhead{Log (L$_{[OIII],corr}$/$\sigma^4$)} & \colhead{$V/V_{max}$} \\
\colhead{erg s$^{-1}$}            &                       & \colhead{erg s$^{3}$km$^{-4}$} & }

\startdata

38.5 - 39.0 & 0.25 & 30.0 - 30.5 & 0.36 \\
39.0 - 39.5 & 0.33 & 30.5 - 31.0 & 0.32 \\
39.5 - 40.0 & 0.38 & 31.0 - 31.5 & 0.37 \\
40.0 - 40.5 & 0.44 & 31.5 - 32.0 & 0.43 \\
40.5 - 41.0 & 0.49 & 32.0 - 32.5 & 0.46 \\
41.0 - 41.5 & 0.51 & 32.5 - 33.0 & 0.50 \\
41.5 - 42.0 & 0.55 & 33.0 - 33.5 & 0.53 \\
$\geq$ 42.0 & 0.60 & $\geq$ 33.5 & 0.59 \\

\enddata

\end{deluxetable}

\section{Results}
\begin{figure}[ht]
\centering
\subfigure[]{\includegraphics[scale=0.28,angle=90]{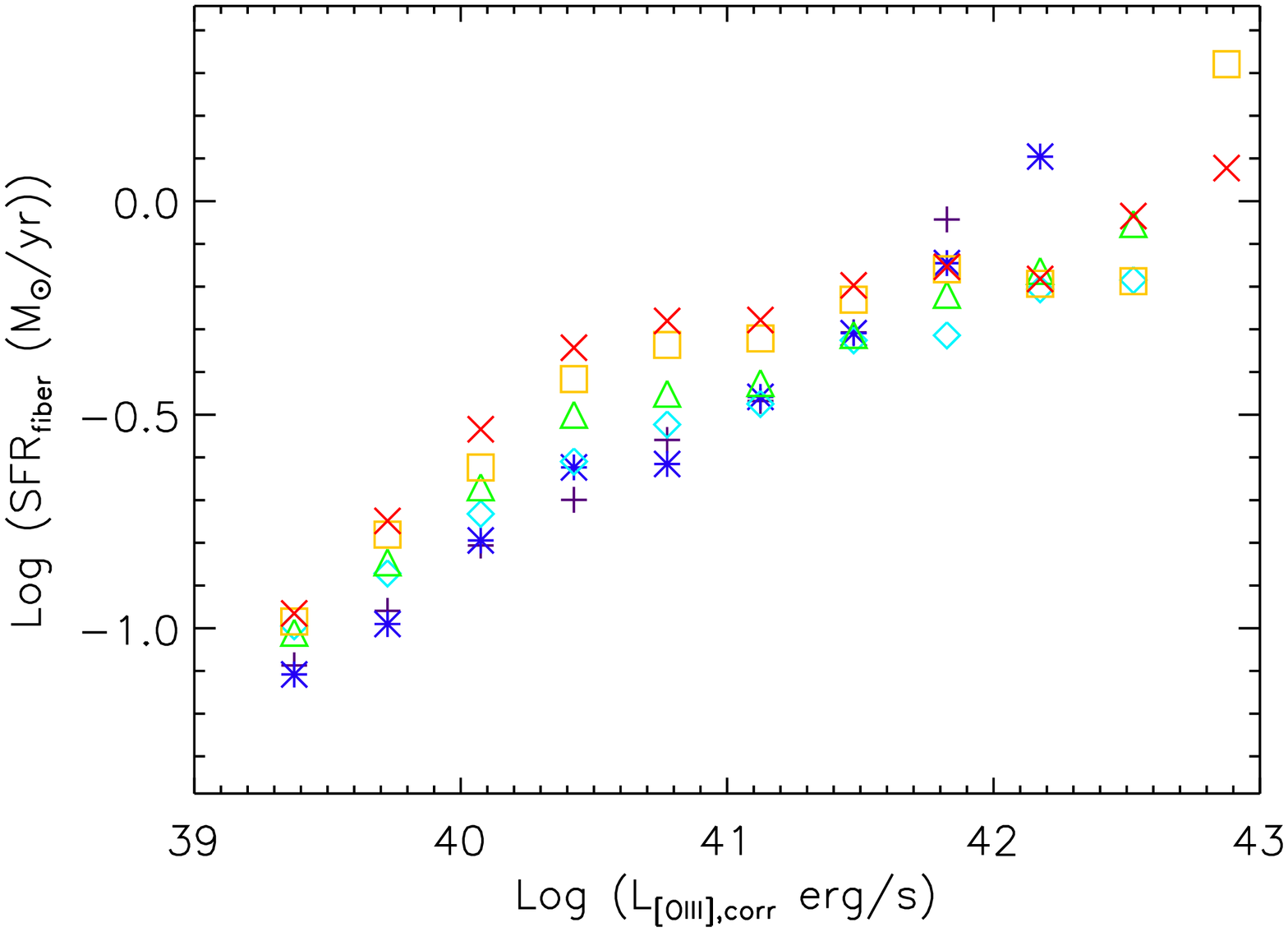}}~
\subfigure[]{\includegraphics[scale=0.28,angle=90]{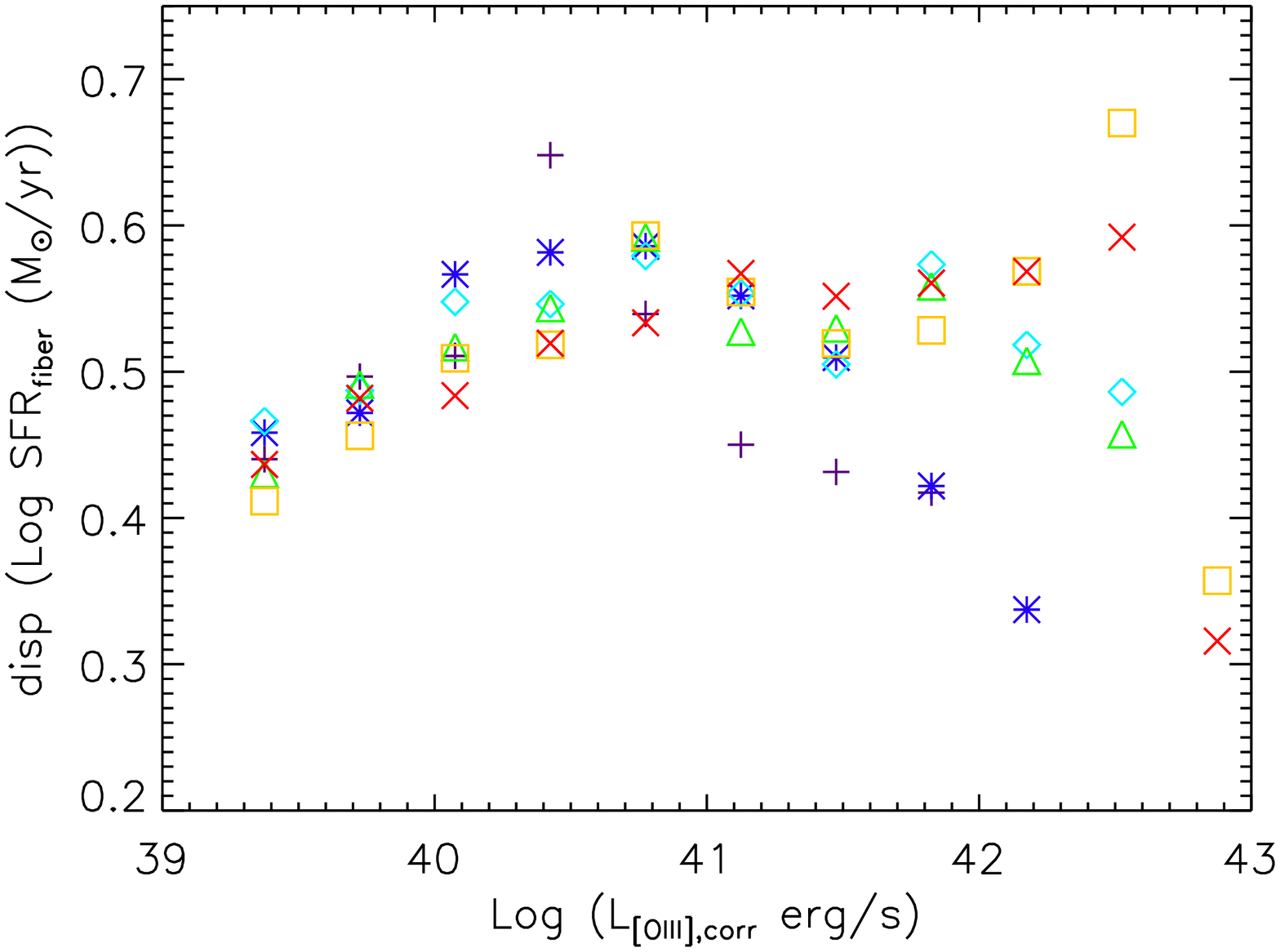}}~
\subfigure[]{\includegraphics[scale=0.3,angle=90]{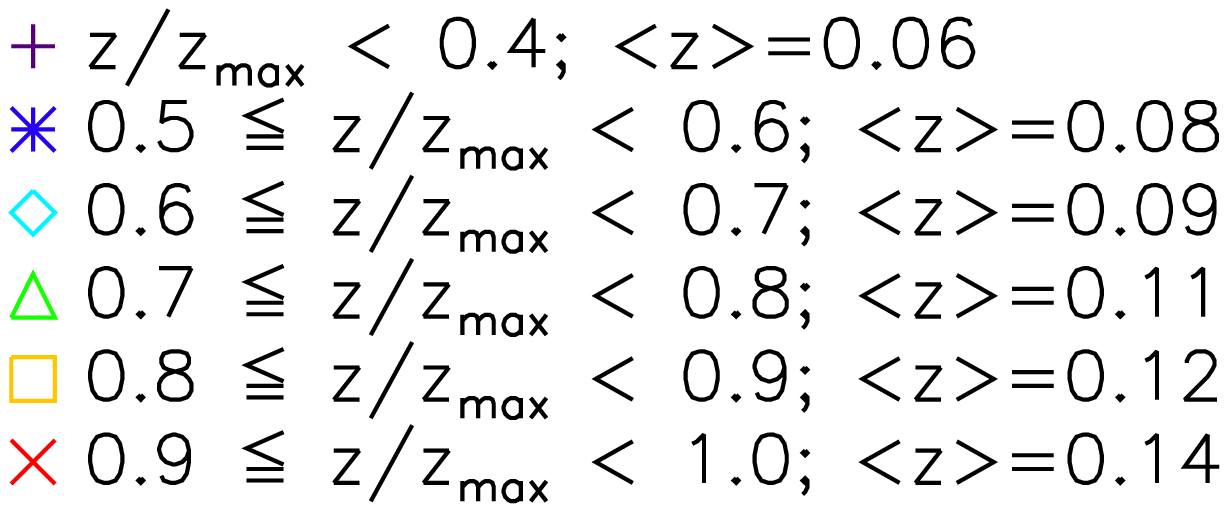}}
\subfigure[]{\includegraphics[scale=0.28,angle=90]{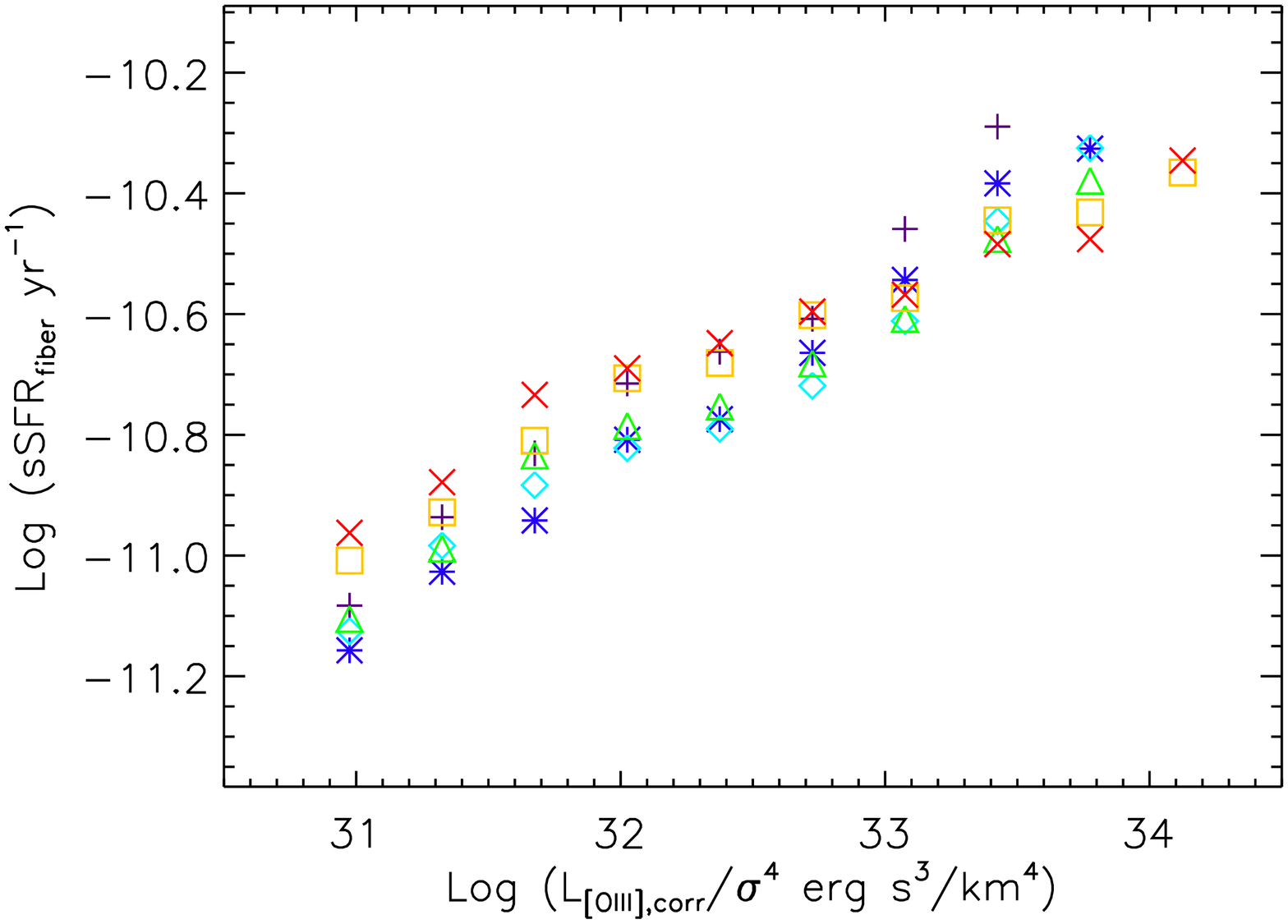}}~
\subfigure[]{\includegraphics[scale=0.28,angle=90]{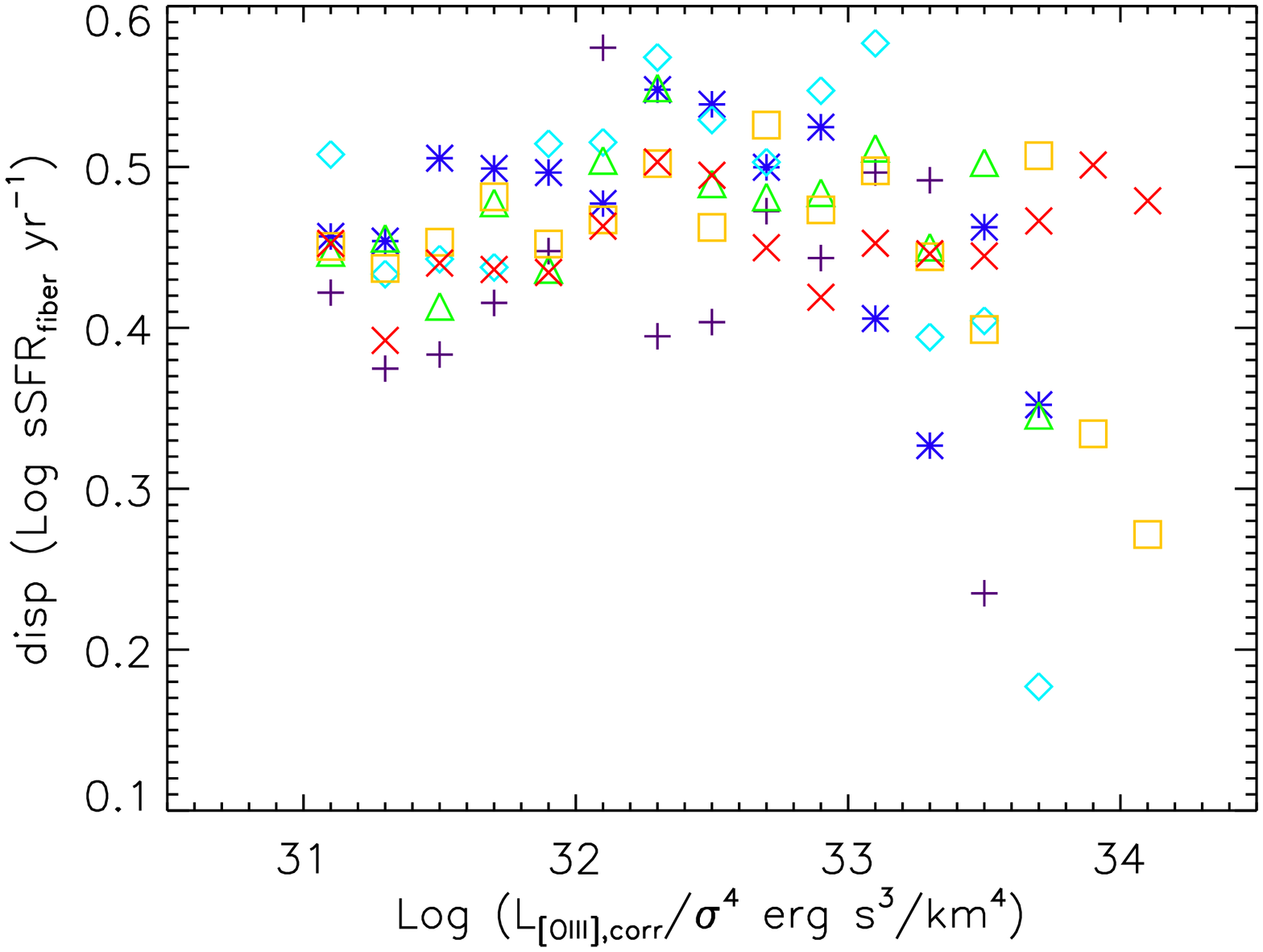}}~
\subfigure[]{\includegraphics[scale=0.3,angle=90]{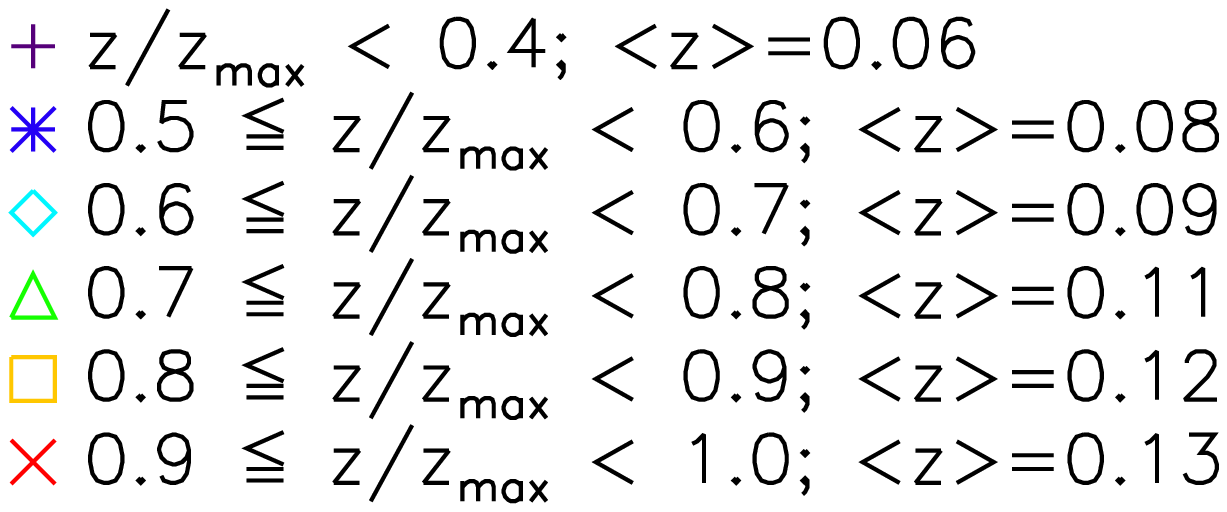}}
\caption[]{\label{sfr_agn}(a) SFR$_{fiber}$ as a function of AGN luminosity and (c) sSFR$_{fiber}$ as a function of Eddington parameter in increasing $z/z_{max}$ bins, (b,d) associated dispersions and (c,e) legends. The legends give the median redshift of the galaxies in each bin: with increasing redshift, the projected physical aperture size grows, encompassing greater amounts of the host galaxy. The relationship between AGN luminosity and star formation shows a dependence on the amount of the host galaxy that is sampled at lower AGN luminosities, but not at higher ones. The slope of the relation between sSFR and Eddington parameter becomes flatter as the aperture size grows, meaning the star formation in more luminous AGN is centrally concentrated.}
\end{figure}

Separating galaxies into bins of $z/z_{max}$, we plot SFR$_{fiber}$ as a function of AGN luminosity (L$_{[OIII],corr}$) and sSFR$_{fiber}$ as a function of the Eddington parameter (L$_{[OIII],corr}$/$\sigma^4$), and the associated dispersions, in Figures \ref{sfr_agn}. Within each z/z$_{max}$ bin, we further bin in AGN luminosity and Eddington parameter, including at least 20 sources in each bin. We then calculate the median SFR and sSFR in these bins. The dispersion ($disp$) is the associated 1$\sigma$ error for a Gaussian distribution, i.e., half the difference between the 83rd and 17th percentiles of the SFR$_{fiber}$ and sSFR$_{fiber}$ distributions in each bin.

As $z/z_{max}$ increases, the 3$^{\prime\prime}$ spectroscopic fiber covers larger physical scales of the host galaxy. The median galaxy redshift ranges from $\sim$0.06 to $\sim$0.13 between the lowest and highest $z/z_{max}$ bin, corresponding to physical aperture radii of 1.7 - 3.5 kpc. As can be seen in Figure \ref{sfr_agn}, the SFR (sSFR) increases slowly but systematically with the AGN luminosity (Eddington ratio), within a given $z/z_{max}$ bin. It is also clear that there are systematic differences between these relations as a function of SDSS aperture size ($z/z_{max}$). At low to moderate AGN luminosities there is a trend for the SFR to grow for a given AGN luminosity as the projected aperture size increases, although this trend is not clear at higher AGN luminosities. Moreover, the slope of the relationship between sSFR and the Eddington-ratio proxy flattens as $z/z_{max}$ increases.
 
To see these trends with distance more clearly, in the left panels of Figure \ref{bin_o3} we plot the star formation rate (top panel) and sSFR (bottom panel) as functions of $z/z_{max}$, and bin the data by the AGN luminosity (top) and Eddington ratio (bottom). These plots demonstrate that the radial variations of SFR and sSFR are dependent on the AGN luminosity and Eddington ratio, respectively. To visualize and quantify how the radial distribution of SFR and sSFR vary as a function of increasing AGN luminosity and Eddington ratio, we have calculated the slope, $a$, of the correlation between
\begin{equation}
(s)SFR \propto (z/z_{max})^a,
\end{equation}
in each AGN luminosity and Eddington parameter bin. For the linear regression fit, we weighted the data points by the error in each bin, taken as $disp/\sqrt{N}$, where $disp$ is the dispersion in the distribution within the bin (half the difference between the 17\% and 83\% percentiles) and $N$ is the number of sources per bin. We note that the lowest $z/z_{max}$ bins have the smallest number of sources, which results in the associated errors being larger.

The results in the right panels of Figure \ref{bin_o3} show the same systematic trend. As the AGN luminosity increases, the increase in star-formation inside the SDSS aperture becomes smaller with greater projected aperture size. At lower AGN luminosities, the relationship between SFR and increasing projected size of the host galaxy is roughly linear, showing that star formation is distributed over the whole range in radial scales we probe in these host galaxies. However, at higher AGN luminosities, star formation is predominantly circumnuclear: we do not see evidence that the SFR increases significantly as the size of region we observe increases from  1.7 to 3.5 kpc (the physical size of the fiber radius at the median redshifts from lowest to highest $z/z_{max}$ bins). Similarly, at high Eddington rates, the negative slope between sSFR and projected aperture size further suggests dominant circumnuclear star formation: as the size of the observed region grows, stellar mass is added, but is not compensated by an increase in star formation.
\section{Discussion}
\begin{figure}[ht]
\centering
\subfigure[]{\includegraphics[scale=0.33,angle=90]{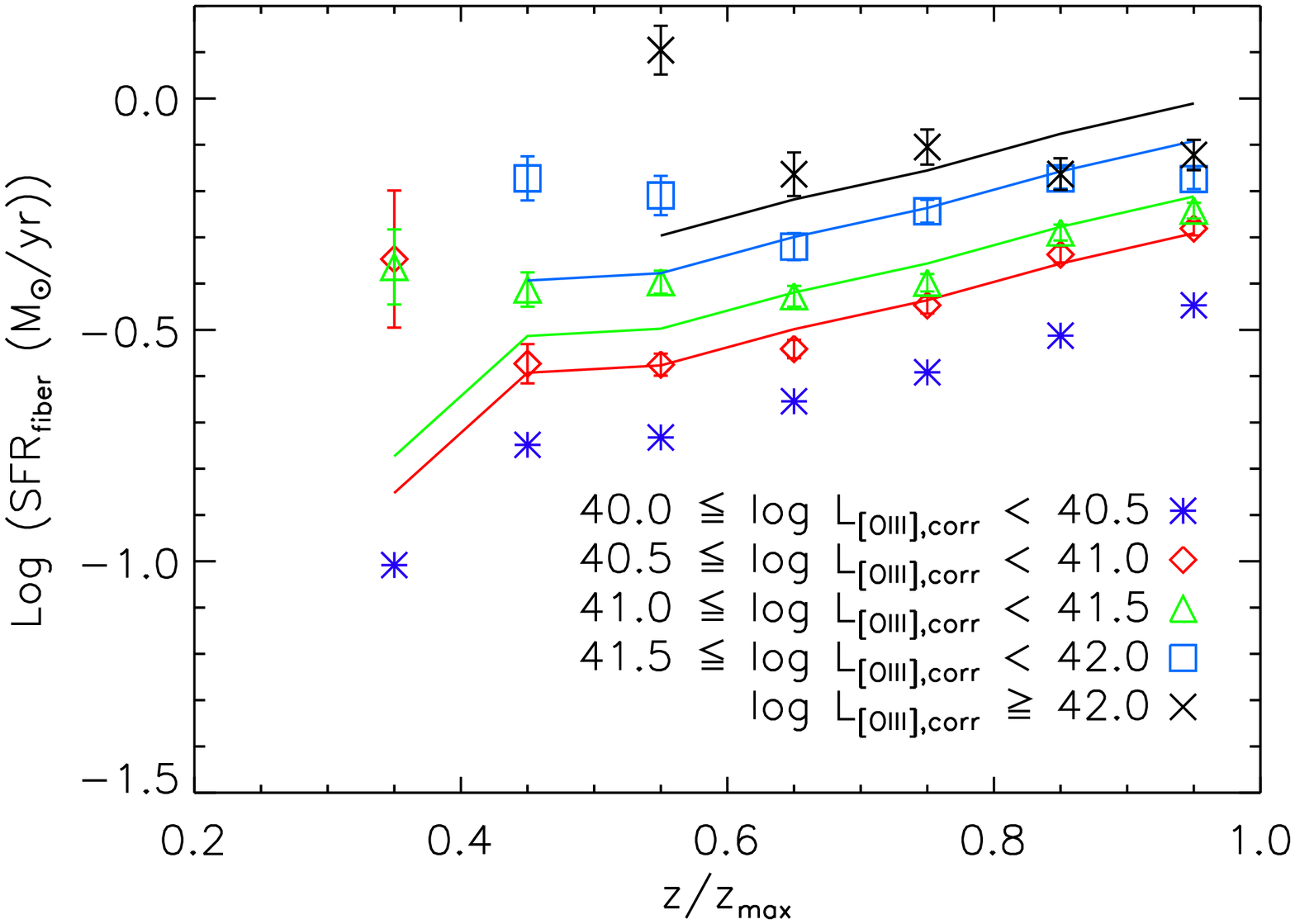}}
\subfigure[]{\includegraphics[scale=0.33,angle=90]{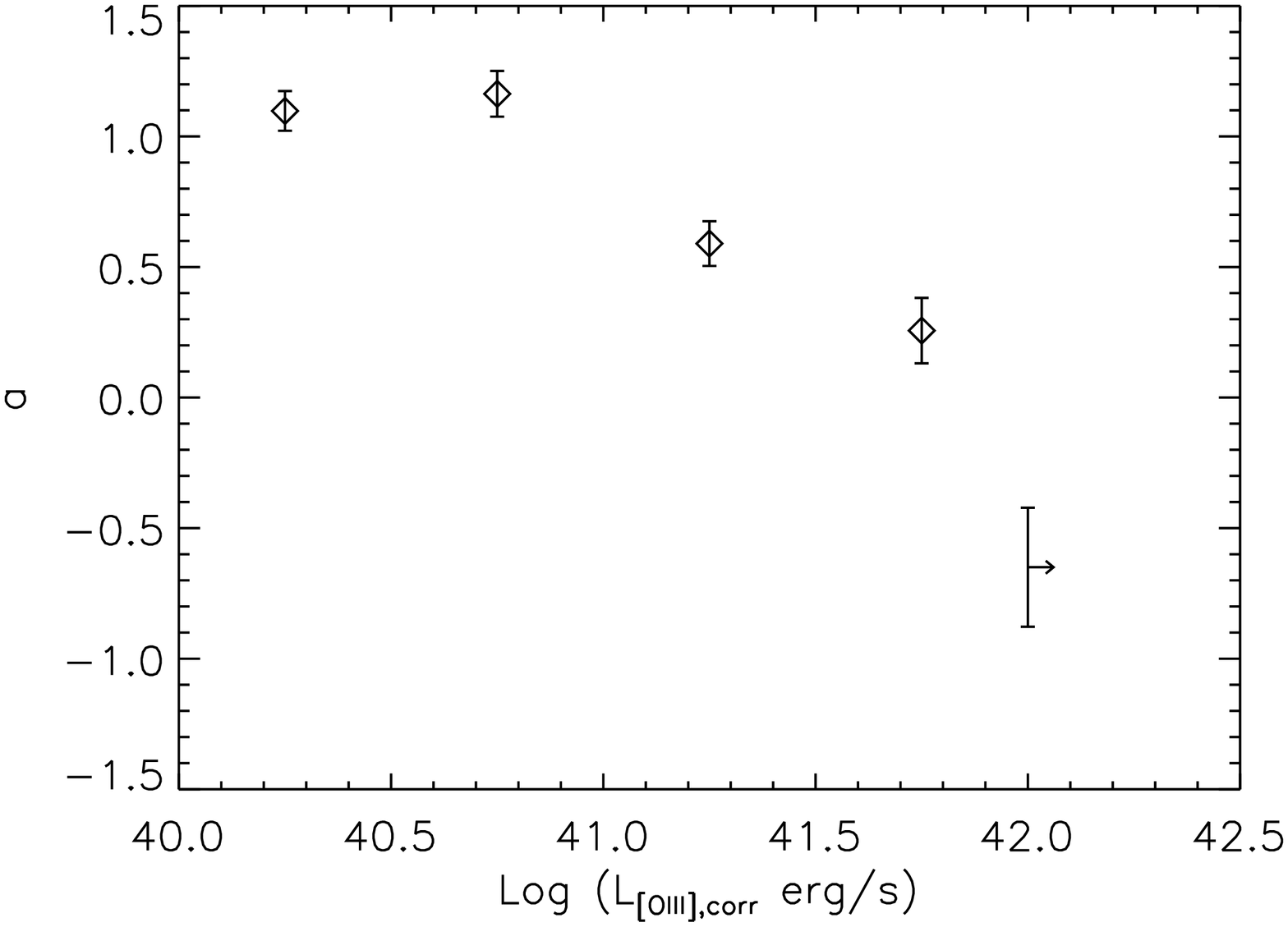}}
\subfigure[]{\includegraphics[scale=0.33,angle=90]{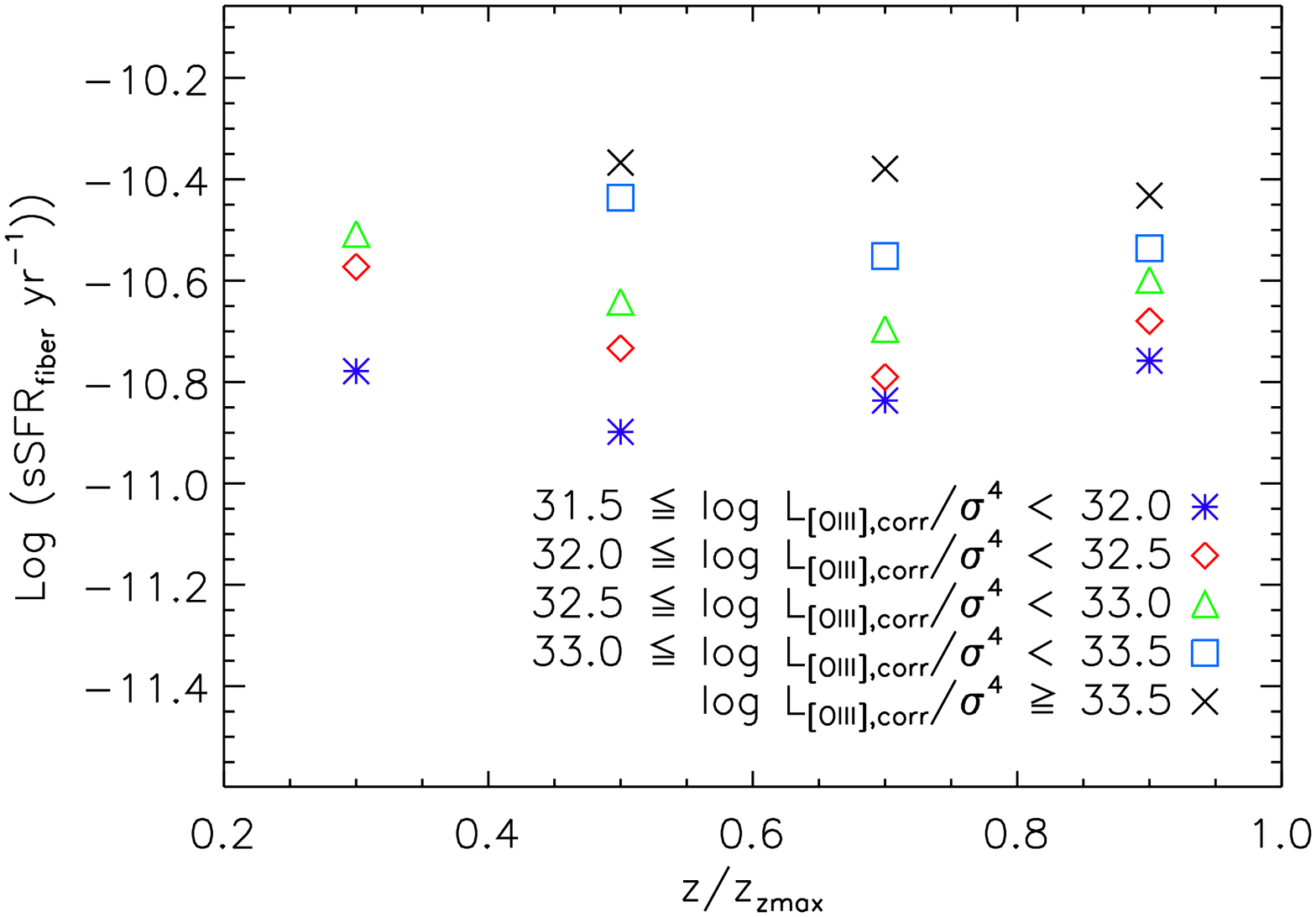}}
\subfigure[]{\includegraphics[scale=0.33,angle=90]{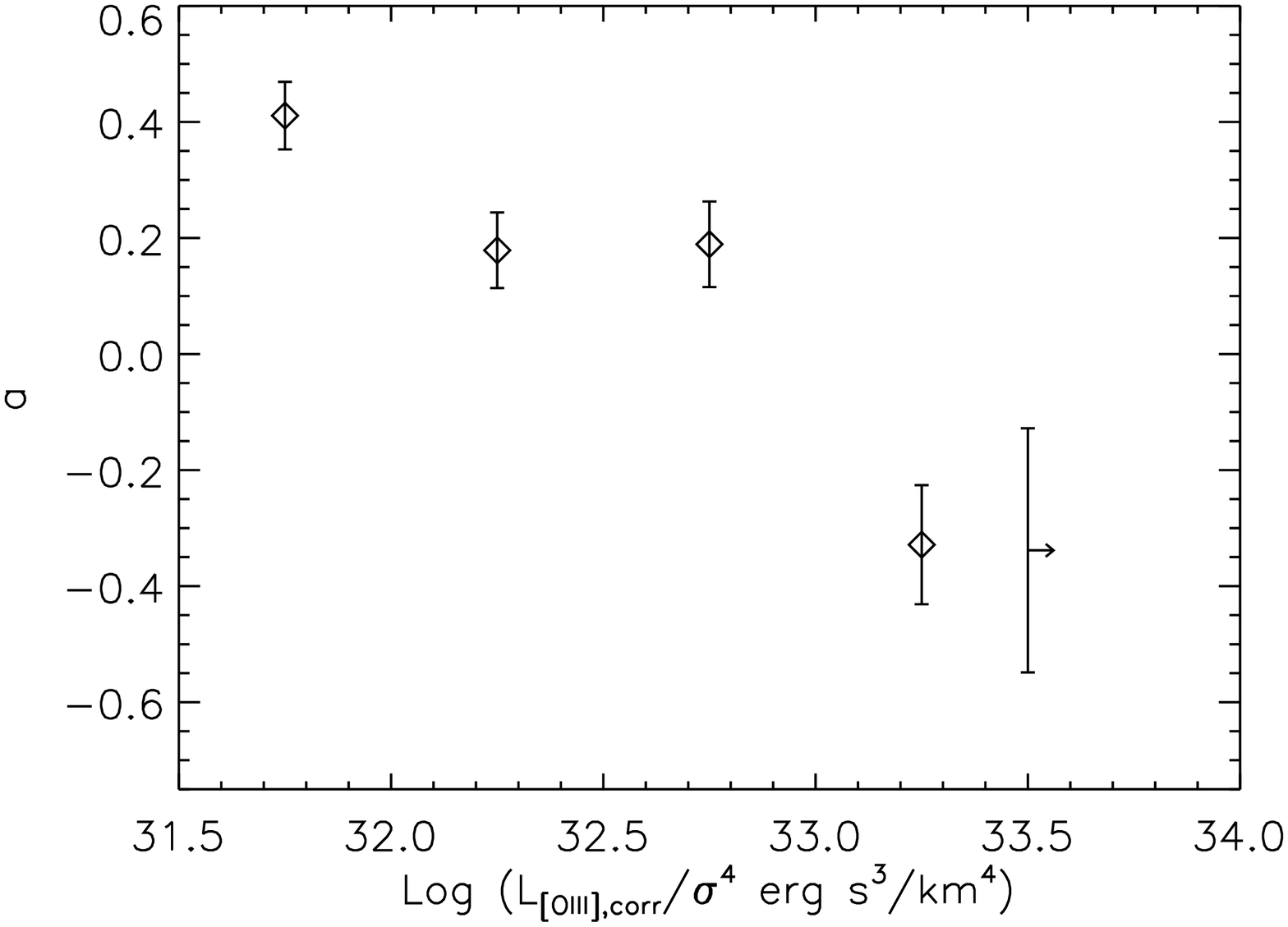}}
\caption[]{\label{bin_o3} (a) Log (SFR$_{fiber}$) as a function of $z/z_{max}$ in increasing L$_{[OIII],corr}$ bins and (b) the best fit slope in each bin.  (c) Log (sSFR$_{fiber}$) as a function of $z/z_{max}$ in increasing L$_{[OIII],corr}/\sigma^4$ bins and (d) the best fit slope in each bin. At higher AGN luminosities and accretion rates, galactic star formation becomes more centrally concentrated. The lines in (a) correspond to the best-fit parameters from SFR$_{fiber}$ = SFR$_{nuclear}$ + SFR$_{disk}$, where SFR$_{nuclear} = \alpha$ (M$_{\sun}$/yr) $\times$(L$_{[OIII],corr}/10^{42}$erg/s)$^\beta$ and SFR$_{disk}$ is SFR$_{fiber}$ associated with the lowest AGN luminosity bin (i.e., 40.0 $<$ log(L$_{[OIII],corr}$ $<$ 40.5 dex). From a reduced $\chi^2$ minimization, we find $\alpha$=0.44$\pm$0.02 and $\beta$=0.36$\pm$0.04.}
\end{figure}

These results indicate that circumnuclear star formation associated with AGN activity (SFR$_{nucleus}$) only dominates over omnipresent disk star formation (SFR$_{disk}$) when the AGN is very luminous. During such a stage, host galaxy star formation will appear to be centrally concentrated. Let us assume
\begin{equation}\label{sfr_eq} 
SFR_{fiber} = SFR_{nucleus} + SFR_{disk},
\end{equation}
The star formation in the nucleus is associated with AGN fueling, parametrized by
\begin{equation}
SFR_{nucleus} = \alpha (M_{\sun}/yr) \times(L_{[OIII],corr}/10^{42} erg/s)^\beta.
\end{equation}
We use SFR$_{fiber}$ from the lowest AGN luminosity bin, 40.0 $<$ log (L$_{[OIII],corr}$) $< 40.5$ dex, as the omnipresent SFR$_{disk}$. We solve equation \ref{sfr_eq} for all log (L$_{[OIII],corr}$) $> 40.5$ dex bins in Figure \ref{bin_o3} (a) simultaneously by minimizing $\chi^2$. Here we take the error on each bin to be $disp$/$\sqrt{N}$, where $disp$ is half the difference between the 83rd and 17th percentile and $N$ refers to the number of sources in the bin. We find $\alpha = 0.44\pm0.02$ and $\beta = 0.36\pm0.04$, and plot these fits in Figure \ref{sfr_eq} (a); the errors represent the 95\% confidence interval, where $\Delta \chi^2$=3.84. These results indicate that star formation associated with SMBH fueling rises sub-linearly with AGN luminosity. The fit could be improved with a more complex model in which both the disk and circumnuclear star formation rates are allowed to have separate functional dependencies on the AGN luminosity, but such complexity is not really warranted. Our simple toy model is only meant to show that the data are consistent with a stronger link between the AGN luminosity and the amount of central star-formation than with larger scale star-formation. 

What can be the physical link between these two processes? Circumnuclear star-formation in luminous AGN can originate from major mergers, as simulations have demonstrated that torques produced by galaxy-galaxy interactions fuel SMBHs while triggering star formation in the galactic center \citep{dimatteo, springel, hopkins05}. Indeed, \citet{treister} have demonstrated that mergers are responsible for triggering the most luminous AGN (L$_{bol} > 10^{45}$ erg s$^{-1}$). However, the majority of the highest luminosity AGN in our sample falls below this bolometric luminosity threshold. Additionally, the lopsidedness of SDSS Sy2 host galaxies within the Eddington range we probe ($\sim10^{-3}$ to one) does not drastically increase, suggesting that mergers are not the primary mechanism for coupling AGN and star formation activity in this sample \citep{Reichard}. Alternatively, \citet{co07} have shown that recycled gas from dying stars near the galactic centers of elliptical galaxies becomes radiatively unstable, causing circumnuclear star formation while also feeding the SMBH. As instabilities caused by bars, bars-in-bars, spiral arms, etc. produce galactic scale star formation while funneling gas to the SMBH \citep{hopkins10}, stellar mass loss seems a more likely mechanism for coupling AGN activity and circumnuclear star formation in moderate luminosity Sy2 galaxies.

\section{Conclusions}

We find that the AGN-star formation connection is largely driven by circumnuclear starburst activity in local Type 2 AGN, consistent with \citet{k07} and \citet{ds}. 

The size-scale of star formation in local active galaxies appears to depend on AGN luminosity: it becomes increasing compact in more luminous AGN. More specifically, the lack of an increase in the amount of SFR as the size of the region probed by the SDSS spectra increases implies that star formation is concentrated in radii below $<$1.7 kpc in these luminous systems. This constrains the mechanism that couples AGN activity and host galaxy star formation, favoring theories that link these two processes at circumnuclear rather than galactic scales (i.e., stellar mass loss rather than instabilities caused from bars, etc.).

Our results are consistent with the hypothesis that circumnuclear star formation is associated with AGN activity and thus dominates over omnipresent galactic disk star formation when the AGN becomes luminous. The dependence of the SFR on SMBH fueling is sub-linear: on radial scales $<$1.7 kpc, $SFR \propto \dot{M}^{0.36}$. This sub-linearity may indicate that accretion is a ``demand'' rather than ``supply'' driven process. Angular momentum transfer through the disk may be relatively inefficient, limiting the accretion rate, regardless of the amount of fuel provided via stellar mass loss. Remaining material could potentially be ejected in winds and/or jets, or perhaps accumulate in the obscuring ``torus.''

\end{document}